\documentclass{emulateapj}
\pdfoutput=1
\usepackage{epstopdf}
\usepackage{natbib}
\usepackage{amsmath}

\DeclareGraphicsExtensions{.jpg,.pdf,.png,.eps,.ps}

\newcommand{\griz}{\protect\hbox{$griz$}}

\newcommand{\sqdeg}{\protect\hbox{deg$^2$} }

\newcommand{\msun}{$M_{\sun}$}
\newcommand{\lsun}{L$_{\sun}$}

\def\lsim{\hbox{\rlap{\raise 0.425ex\hbox{$<$}}\lower 0.65ex\hbox{$\sim$}}}
\def\gsim{\hbox{\rlap{\raise 0.425ex\hbox{$>$}}\lower 0.65ex\hbox{$\sim$}}}

\def\arcsec{\hbox{$^{\prime\prime}$}}
\def\arcdeg{\mbox{$^\circ$}}

\newcommand{\cluster}{SPT-CL~J0205-5829}
\newcommand{\uber}{SPT-CL~J2106-5844}
\newcommand{\rarestcluster}{SPT-CL~J0102-4915}
\newcommand{\zp}{$z = 1.30 \pm 0.12$}
\newcommand{\zpnoe}{$z = 1.30$}
\newcommand{\zs}{$z = 1.322^{+0.001}_{-0.002}$}
\newcommand{\zbcg}{$z=1.3218 \pm  0.0005$}

\newcommand{\zsnoe}{$z = 1.322$}

\newcommand{\xtemp}{$T_X = 8.7^{+1.0}_{-0.8}\,$keV}

\newcommand{\xlum}{$L_{X} ({\rm 0.5 - 2.0~keV}) = (3.91 \pm 0.05) \times 10^{44}$~erg~s$^{-1}$}
\newcommand{\uk}{\mu \mathrm{K}}

\newcommand{\xtempt}{\ensuremath{8.7^{+1.0}_{-0.8}}}
\newcommand{\xtemptrosati}{\ensuremath{8.6^{+1.3}_{-1.2}}}

\newcommand{\xcts}{\ensuremath{5500}}
\newcommand{\massszfive}{\ensuremath{4.8 \pm 1.0}}
\newcommand{\masssztwo}{\ensuremath{8.7 \pm 1.8}}
\newcommand{\masstxtfive}{\ensuremath{5.2 \pm 1.3}}

\newcommand{\masstxtmeas}{\ensuremath{(5.2 \pm 1.3) \times 10^{14}}}
\newcommand{\massbestshortfive}{\ensuremath{4.8 \pm 0.8}}
\newcommand{\massbestfive}{\ensuremath{(4.8 \pm 0.8) \times 10^{14}}}

\newcommand{\massbesttwo}{\ensuremath{(8.8 \pm 1.4) \times 10^{14}}}
\newcommand{\rd}{\mbox{$r_{500}$}}

\newcommand{\md}{\mbox{$M_{500}$}}

\shorttitle{\cluster\ at \lowercase{$z =  1.32$}}
\shortauthors{Stalder et~al.}

\begin{document}

\title{\cluster: a \lowercase{$z = 1.32$} Evolved Massive Galaxy Cluster in the South Pole Telescope Sunyaev-Zel'dovich Effect Survey}

\altaffiltext{\CfA}{Harvard-Smithsonian Center for Astrophysics,
60 Garden Street, Cambridge, MA 02138}
\altaffiltext{\Harvard}{Department of Physics, Harvard University, 17 Oxford Street, Cambridge, MA 02138}
\altaffiltext{\Munich}{Department of Physics,
Ludwig-Maximilians-Universit\"{a}t,
Scheinerstr.\ 1, 81679 M\"{u}nchen, Germany}
\altaffiltext{\Miss}{Department of Physics, University of Missouri, 5110 Rockhill Road, Kansas City, MO 64110}
\altaffiltext{\UChicago}{University of Chicago,
5640 South Ellis Avenue, Chicago, IL 60637}
\altaffiltext{\MIT}{MIT Kavli Institute for Astrophysics and Space
Research, Massachusetts Institute of Technology, 77 Massachusetts Avenue,
Cambridge, MA 02139}
\altaffiltext{\NCSA}{National Center for Supercomputing Applications,
University of Illinois, 1205 West Clark Street, Urbanan, IL 61801}
\altaffiltext{\ExcellenceCluster}{Excellence Cluster Universe,
Boltzmannstr.\ 2, 85748 Garching, Germany}
\altaffiltext{\KICPChicago}{Kavli Institute for Cosmological Physics,
University of Chicago,
5640 South Ellis Avenue, Chicago, IL 60637}
\altaffiltext{\EFIChicago}{Enrico Fermi Institute,
University of Chicago,
5640 South Ellis Avenue, Chicago, IL 60637}
\altaffiltext{\PhysicsUChicago}{Department of Physics,
University of Chicago,
5640 South Ellis Avenue, Chicago, IL 60637}
\altaffiltext{\AAUChicago}{Department of Astronomy and Astrophysics,
University of Chicago,
5640 South Ellis Avenue, Chicago, IL 60637}
\altaffiltext{\ANL}{Argonne National Laboratory, 9700 S. Cass Avenue, Argonne, IL, USA 60439}
\altaffiltext{\NIST}{NIST Quantum Devices Group, 325 Broadway Mailcode 817.03, Boulder, CO, USA 80305}
\altaffiltext{\PUC}{Departamento de Astronomia y Astrosifica, Pontificia Universidad Catolica,
Chile}
\altaffiltext{\McGill}{Department of Physics,
McGill University,
3600 Rue University, Montreal, Quebec H3A 2T8, Canada}
\altaffiltext{\Berkeley}{Department of Physics,
University of California, Berkeley, CA 94720}
\altaffiltext{\UFlorida}{Department of Astronomy, University of Florida, Gainesville, FL 32611}
\altaffiltext{\Colorado}{Department of Astrophysical and Planetary Sciences and Department of Physics,
University of Colorado,
Boulder, CO 80309}
\altaffiltext{\NASA}{Department of Space Science, VP62,
NASA Marshall Space Flight Center,
Huntsville, AL 35812}
\altaffiltext{\Davis}{Department of Physics, 
University of California, One Shields Avenue, Davis, CA 95616}
\altaffiltext{\LBNL}{Physics Division,
Lawrence Berkeley National Laboratory,
Berkeley, CA 94720}
\altaffiltext{\Caltech}{California Institute of Technology, 1200 E. California Blvd., Pasadena, CA 91125}
\altaffiltext{\Arizona}{Steward Observatory, University of Arizona, 933 North Cherry Avenue, Tucson, AZ 85721}
\altaffiltext{\Michigan}{Department of Physics, University of Michigan, 450 Church Street, Ann  
Arbor, MI, 48109}
\altaffiltext{\MPE}{Max-Planck-Institut f\"{u}r extraterrestrische Physik,
Giessenbachstr.\ 85748 Garching, Germany}
\altaffiltext{\CaseWestern}{Physics Department, Center for Education and Research in Cosmology 
and Astrophysics, 
Case Western Reserve University,
Cleveland, OH 44106}
\altaffiltext{\Minnesota}{Physics Department, University of Minnesota, 116 Church Street S.E., Minneapolis, MN 55455}
\altaffiltext{\STScI}{Space Telescope Science Institute, 3700 San Martin
Dr., Baltimore, MD 21218}
\altaffiltext{\SAIC}{Liberal Arts Department, 
School of the Art Institute of Chicago, 
112 S Michigan Ave, Chicago, IL 60603}
\altaffiltext{\Yale}{Department of Physics, Yale University, P.O. Box 208210, New Haven,
CT 06520-8120}
\altaffiltext{\LLNL}{Institute of Geophysics and Planetary Physics, Lawrence
Livermore National Laboratory, Livermore, CA 94551}
\altaffiltext{\BCCP}{Berkeley Center for Cosmological Physics,
Department of Physics, University of California, and Lawrence Berkeley
National Labs, Berkeley, CA 94720}

\def\CfA{1}
\def\Harvard{2}
\def\Munich{3}
\def\Miss{4}
\def\UChicago{5}
\def\MIT{6}
\def\NCSA{7}
\def\ExcellenceCluster{8}
\def\KICPChicago{9}
\def\EFIChicago{10}
\def\PhysicsUChicago{11}
\def\AAUChicago{12}
\def\ANL{13}
\def\NIST{14}
\def\PUC{15}
\def\McGill{16}
\def\Berkeley{17}
\def\UFlorida{18}
\def\Colorado{19}
\def\NASA{20}
\def\Davis{21}
\def\LBNL{22}
\def\Caltech{23}
\def\Arizona{24}
\def\Michigan{25}
\def\MPE{26}
\def\CaseWestern{27}
\def\Minnesota{28}
\def\STScI{29}
\def\SAIC{30}
\def\Yale{31}
\def\LLNL{32}
\def\BCCP{33}

\author{B.~Stalder\altaffilmark{\CfA},
J.~Ruel\altaffilmark{\Harvard},
R.~\v{S}uhada\altaffilmark{\Munich},
M.~Brodwin\altaffilmark{\Miss},
K.~A.~Aird\altaffilmark{\UChicago},
K.~Andersson\altaffilmark{\Munich,\MIT},
R.~Armstrong\altaffilmark{\NCSA},
M.~L.~N.~Ashby\altaffilmark{\CfA},
M.~Bautz\altaffilmark{\MIT},
M.~Bayliss\altaffilmark{\Harvard}, 
G.~Bazin\altaffilmark{\Munich,\ExcellenceCluster},
B.~A.~Benson\altaffilmark{\KICPChicago,\EFIChicago},
L.~E.~Bleem\altaffilmark{\KICPChicago,\PhysicsUChicago},
J.~E.~Carlstrom\altaffilmark{\KICPChicago,\EFIChicago,\PhysicsUChicago,\AAUChicago,\ANL}, 
C.~L.~Chang\altaffilmark{\KICPChicago,\EFIChicago,\ANL}, 
H.~M. Cho\altaffilmark{\NIST}, 
A.~Clocchiatti\altaffilmark{\PUC},
T.~M.~Crawford\altaffilmark{\KICPChicago,\AAUChicago},
A.~T.~Crites\altaffilmark{\KICPChicago,\AAUChicago},
T.~de~Haan\altaffilmark{\McGill},
S.~Desai\altaffilmark{\Munich,\ExcellenceCluster},
M.~A.~Dobbs\altaffilmark{\McGill},
J.~P.~Dudley\altaffilmark{\McGill},
R.~J.~Foley\altaffilmark{\CfA}, 
W.~R.~Forman\altaffilmark{\CfA},
E.~M.~George\altaffilmark{\Berkeley},
D.~Gettings\altaffilmark{\UFlorida},
M.~D.~Gladders\altaffilmark{\KICPChicago,\AAUChicago},
A.~H.~Gonzalez\altaffilmark{\UFlorida},
N.~W.~Halverson\altaffilmark{\Colorado},
N.~L.~Harrington\altaffilmark{\Berkeley},
F.~W.~High\altaffilmark{\KICPChicago,\AAUChicago}, 
G.~P.~Holder\altaffilmark{\McGill},
W.~L.~Holzapfel\altaffilmark{\Berkeley},
S.~Hoover\altaffilmark{\KICPChicago,\EFIChicago},
J.~D.~Hrubes\altaffilmark{\UChicago},
C.~Jones\altaffilmark{\CfA},
M.~Joy\altaffilmark{\NASA},
R.~Keisler\altaffilmark{\KICPChicago,\PhysicsUChicago},
L.~Knox\altaffilmark{\Davis},
A.~T.~Lee\altaffilmark{\Berkeley,\LBNL},
E.~M.~Leitch\altaffilmark{\KICPChicago,\AAUChicago},
J.~Liu\altaffilmark{\Munich,\ExcellenceCluster},
M.~Lueker\altaffilmark{\Berkeley,\Caltech},
D.~Luong-Van\altaffilmark{\UChicago},
A.~Mantz\altaffilmark{\KICPChicago},
D.~P.~Marrone\altaffilmark{\Arizona},
M.~McDonald\altaffilmark{\MIT},
J.~J.~McMahon\altaffilmark{\KICPChicago,\EFIChicago,\Michigan},
J.~Mehl\altaffilmark{\KICPChicago,\AAUChicago},
S.~S.~Meyer\altaffilmark{\KICPChicago,\PhysicsUChicago,\EFIChicago,\AAUChicago},
L.~Mocanu\altaffilmark{\KICPChicago,\AAUChicago},
J.~J.~Mohr\altaffilmark{\Munich,\ExcellenceCluster,\MPE},
T.~E.~Montroy\altaffilmark{\CaseWestern},
S.~S.~Murray\altaffilmark{\CfA},
T.~Natoli\altaffilmark{\KICPChicago,\PhysicsUChicago},
D.~Nurgaliev\altaffilmark{\Harvard}, 
S.~Padin\altaffilmark{\KICPChicago,\AAUChicago,\Caltech},
T.~Plagge\altaffilmark{\KICPChicago,\AAUChicago},
C.~Pryke\altaffilmark{\Minnesota}, 
C.~L.~Reichardt\altaffilmark{\Berkeley},
A.~Rest\altaffilmark{\STScI},
J.~E.~Ruhl\altaffilmark{\CaseWestern}, 
B.~R.~Saliwanchik\altaffilmark{\CaseWestern}, 
A.~Saro\altaffilmark{\Munich},
J.~T.~Sayre\altaffilmark{\CaseWestern}, 
K.~K.~Schaffer\altaffilmark{\KICPChicago,\EFIChicago,\SAIC}, 
L.~Shaw\altaffilmark{\McGill,\Yale},
E.~Shirokoff\altaffilmark{\Berkeley,\Caltech}, 
J.~Song\altaffilmark{\Michigan},
H.~G.~Spieler\altaffilmark{\LBNL},
S.~A.~Stanford\altaffilmark{\Davis,\LLNL},
Z.~Staniszewski\altaffilmark{\CaseWestern},
A.~A.~Stark\altaffilmark{\CfA}, 
K.~Story\altaffilmark{\KICPChicago,\PhysicsUChicago},
C.~W.~Stubbs\altaffilmark{\CfA,\Harvard}, 
A.~van~Engelen\altaffilmark{\McGill},
K.~Vanderlinde\altaffilmark{\McGill},
J.~D.~Vieira\altaffilmark{\KICPChicago,\PhysicsUChicago,\Caltech},
A. Vikhlinin\altaffilmark{\CfA},
R.~Williamson\altaffilmark{\KICPChicago,\AAUChicago}, 
O.~Zahn\altaffilmark{\Berkeley,\BCCP},
and
A.~Zenteno\altaffilmark{\Munich,\ExcellenceCluster}
}
\email{bstalder@cfa.harvard.edu}
\slugcomment{Submitted to \apj}

\begin{abstract}

The galaxy cluster \cluster\ currently has the highest spectroscopically-confirmed redshift, \lowercase{\zsnoe}, in the South Pole Telescope Sunyaev-Zel'dovich (SPT-SZ) survey.  XMM-\emph{Newton} observations measure a core-excluded temperature of \xtemp\ producing a mass estimate that is consistent with the Sunyaev-Zel'dovich derived mass.  The combined SZ and X-ray mass estimate of $M_{500}$=\massbestfive$h_{70}^{-1}$\msun\ makes it the most massive known SZ-selected galaxy cluster at $z > 1.2$ and the second most massive at $z > 1$.  Using optical and infrared observations, we find that the brightest galaxies in \cluster\ are already well evolved by the time the universe was $<$5 Gyr old, with stellar population ages $\gsim$3 Gyr, and low rates of star formation ($<$0.5\msun/yr).  We find that, despite the high redshift and mass, the existence of \cluster\ is not surprising given a flat $\Lambda$CDM cosmology with Gaussian initial perturbations.  The a priori chance of finding a cluster of similar rarity (or rarer) in a survey the size of the 2500\,{\rm deg}$^2$ SPT-SZ survey is 69\%.


\end{abstract}



\setcounter{footnote}{1}

\section{Introduction}\label{s:intro}

The South Pole Telescope \citep[SPT;][]{carlstrom11} has recently completed a survey
designed to discover all massive galaxy clusters within a 2500~\sqdeg region of the southern sky.
High redshift galaxy clusters are valuable as probes of the
initial conditions of the universe, particularly the distribution of
matter at early epochs.  Since galaxy clusters are the most massive
collapsed systems, 
their abundance is sensitive to the properties of the early universe including Gaussianity
around the peak of the matter density field \citep[e.g.,][]{lucchin88,
colafrancesco89, mortonson10} and the nature of inflationary models.
In addition to cosmology, the
constituent galaxies of these clusters, which have essentially co-evolving
star formation histories, are useful for studying galaxy
formation and evolution.

The SPT-SZ survey finds clusters via the Sunyaev-Zel'dovich \citep[SZ;][]{sunyaev72} effect.
The vast majority of baryonic
mass of a galaxy cluster is in the form of diffuse, ionized gas, known as the intracluster medium (ICM),
unassociated with any particular galaxy.
Photons from the cosmic microwave
background (CMB) are Compton scattered by the free electrons in this ionized
gas.  The scattered photons gain energy on average leading to a spectral distortion of the observed CMB known as the
thermal SZ effect.  The surface brightness of the SZ effect is independent of
the distance to the cluster because the SZ effect depends solely on the
line-of-sight integral of thermal pressure of the ionized gas.
Therefore the total SZ flux is a measure of the total thermal
energy in the gas, which is tightly correlated to the cluster mass.
This makes SZ surveys an efficient means for finding high mass
clusters at all redshifts \citep[e.g.,][]{carlstrom02}.

Observations of fine scale CMB anisotropy with the SPT,
Planck Satellite \citep{planck11-5.1a},
and Atacama Cosmology Telescope \citep[ACT;][]{marriage11b}
have recently been used to detect massive clusters
in large surveys of the sky.
The progress of the cluster survey by SPT is reported
by \citet{staniszewski09,vanderlinde10,williamson11,reichardt12c}
where the details of the survey strategy, data reduction, and
cosmological analysis are also presented.
The SPT has now completed a survey of 2500~\sqdeg in the southern hemisphere
in three millimeter-wavelength bands.
The SPT-SZ survey is essentially complete for clusters with a mass of
$M_{500}\gtrsim 5 \times 10^{14}$~$h_{70}^{-1}$\msun\ at $z > 0.3$ and 
$M_{500}\gtrsim 3 \times 10^{14}$~$h_{70}^{-1}$\msun\ at $z > 1$ .

The SPT-SZ survey has discovered several
galaxy clusters that have been spectroscopically confirmed at $z > 1 $
\citep{brodwin10,foley11}.  Other groups are also discovering high
redshift clusters through X-ray
\citep[e.g.,][]{rosati04, mullis05, stanford06, rosati09,henry10}, or
infrared imaging \citep[e.g.,][]{stanford05,
brodwin06, eisenhardt08, muzzin09, wilson09, papovich10,brodwin11,stanford12}.
However, these techniques do not have the benefit of simultaneously providing the nearly
redshift-independent mass selection and wide area coverage of the SZ technique.

The rich and dense environments of galaxy clusters can also be exploited to study galaxy evolution.  They provide a simple way of identifying large populations of galaxies that have similar formation histories.  As the redshift of the clusters in the sample increases, earlier phases in the process of galaxy evolution process are observed.   
With these observations, scenarios of galaxy formation can be tested with less temporal extrapolation.
This also perhaps has ramifications for the hierarchical formation scenario, as the ages of the bright elliptical galaxies may be related to the epoch of the final assembly of the cluster as suggested by semi-analytical simulations such as \citet{dubinski98} and \citet{boylan09}.  Recent observations of high redshift dense environments point to a period of prodigious star formation \citep{papovich12,snyder12} at $z > 1.3$, after which the bulk of the galaxy stellar mass build up likely arises from dry mergers (without significant star formation).  However, studies of lower redshift elliptical galaxies suggest that stellar age is minimally affected by environment densities, e.g. \citet{thomas10a}.  A larger data set at high redshift is probably required to reconcile these.


\cluster\ was first identified as a cluster in \citet[][R12]{reichardt12c},
which describes a catalog of 224 cluster
candidates discovered in the first $720\,{\rm deg}^2$ of the 
$2500\,{\rm deg}^2$ SPT-SZ survey.  \cluster\ was detected
with a signal-to-noise ratio (S/N) of 10.5 in the SPT data.  
Initial deep optical
follow-up observations showed no obvious overdensity of galaxies in
\griz\ images, but additional infrared and {\it Spitzer} photometry
confirmed the presence of extremely red clustered galaxies consistent with
a redshift $z > 1.3$.  Optical spectroscopy of member galaxies confirmed
that the cluster is at \zsnoe.  X-ray observations with 
XMM-\emph{Newton} revealed a luminous and
extended X-ray source.    Although \cluster\ is not
the most massive SPT cluster, it is
the highest-redshift SPT cluster that has been confirmed by spectroscopy
to date, and potentially the most massive galaxy cluster known at redshift $z > 1.2$
(previous to this was XMM2235 from \citet{rosati09} at $z$=1.39), and second most
massive at $z > 1$ (the most massive being SPT-CL J2106-5844 from \citet{foley11} at $z$=1.13).

We present our initial detection and follow-up observations of
\cluster\ in Section~\ref{s:obs}.  In Section~\ref{s:results}, we show
that \cluster\ is a massive high redshift galaxy cluster 
with a population of normal passively evolving galaxies.  We then briefly
discuss the implications of the existence of such a
massive, evolved cluster at $z > 1.3$ in Section~\ref{s:discussion}.  We
summarize and conclude in Section~\ref{s:conc}.  Except where otherwise stated,
we assume a flat $\Lambda$CDM cosmology with $\Omega_M$=0.3 and $h_0$=0.7 throughout 
this paper.
$M_{500}$ masses are defined as the mass enclosed
in a spherical region which has a density 500 times the critical density of the universe.
At $z$=1.322, 1 Mpc subtends 2.0 arcminutes and the age of the universe is
4.66 Gyr.


\section{Observations, Data Reduction, \& Initial Findings}\label{s:obs}

\subsection{Millimeter Observations by The South Pole Telescope}

\cluster\ was initially discovered in the SPT-SZ survey and reported in R12, as part of 
the cluster catalog identified from the 720~\sqdeg surveyed during the 2008-2009 SPT observing seasons.  
The survey strategy
and data analysis are detailed in the previous SPT-SZ
survey papers \citet{staniszewski09}, \citet{vanderlinde10},
\citet[][W11]{williamson11}, and R12.  The SPT-SZ survey was completed 
in November 2011, and covers an area of 2500~\sqdeg in three frequency bands
at 95, 150, and 220 GHz.  

\begin{figure*}
\begin{center}
\plotone{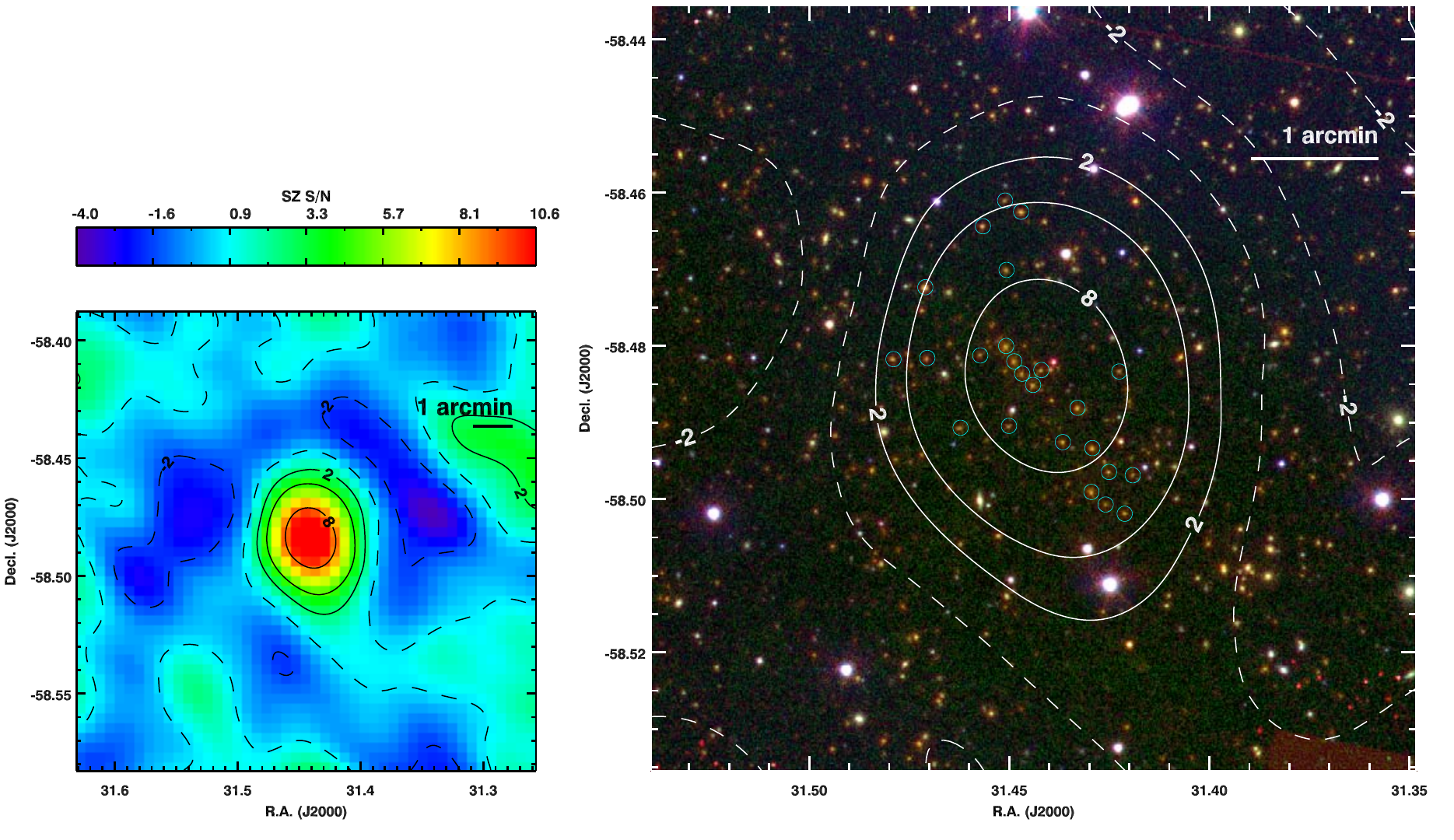}
\caption{
(Left) The filtered SPT-SZ significance map of \cluster.
The negative trough surrounding the cluster is an artifact of the filtering of the time ordered data and maps. (Right) Color image from IMACS i, NEWFIRM K$_S$, Spitzer/IRAC [3.6], with SPT-SZ contours overlayed in white and [3.6]-[4.5] color-selected galaxies indicated in cyan.
}\label{f:sz}
\end{center}
\end{figure*}

As described in R12, cluster candidates were identified using a 
multi-band matched-filter approach, similar to that
first described by \citet{melin06}.  The significance of
a cluster detection (maximized across spatial filter scales and
position in map), $\xi$, is used to identify cluster candidates.
For the survey field containing \cluster, only the 95 and 150 GHz data was used, 
the SPT maps have noise levels of 45 and 16~$\uk$-arcmin in CMB temperature units at
 95 and 150~GHz, respectively. 
In this data, \cluster\ was detected with $\xi = 10.5$ and is among the 5\% most significant
detections in the R12 catalog.  An image of the filtered SPT map is shown in Figure \ref{f:sz}.

\subsection{Optical and Infrared Imaging}

We obtained \griz\ imaging using the MOSAIC2 imager on the CTIO $4\,\mathrm{m}$
Blanco telescope on UT 18, 25 July 2010 and UT 4 July 2011 with mediocre to bad seeing (1.1 to 2.2'') 
and occasional light clouds.  
Total integration times were 300, 300, 2350, and 1050 seconds to 10-$\sigma$ point source depths of 
23.8, 23.2, 22.2, and 21.1 AB magnitudes in g, r, i, and z, respectively.
We also acquired 1800 seconds of deep i-band imaging of \cluster\ on UT 31 January 2011 with
 the Inamori Magellan Areal Camera and Spectrograph
\citep[IMACS;][]{dressler06} on the Baade Magellan $6.5\,\mathrm{m}$
telescope to 24.0 AB magnitude depth in mediocre seeing (1.2 to 1.5'').  The observation strategy and reduction procedure is
described in \citet[][H10]{high10}, W11, and \citet[][S12]{song12b} using the 
PHOTPIPE pipeline \citep{rest05a}.

\cluster\ was also observed with the NEWFIRM imager  \citep{autry03} 
at the CTIO $4\,\mathrm{m}$ Blanco telescope on UT 6 November 2010.
Data were obtained in the $K_{s}$ filter under photometric conditions with a 10-$\sigma$ point source depth of 19.1 Vega magnitudes.
At each dither position, six frames with 10~s exposure times were
coadded
at 18 random positions providing a total exposure time of
1080~s.  NEWFIRM data were reduced using the \texttt{FATBOY} pipeline, originally
developed for the FLAMINGOS-2 instrument, and modified to work with
NEWFIRM data in support of the Infrared Bootes Imaging Survey
(A. Gonzalez, private communication).  Individual processed frames are combined using
SCAMP and SWARP \citep{bertin02}, and photometry is calibrated to
2MASS \citep{skrutskie06}.  The final image has a FWHM
of 0.96\arcsec\!\!.

\begin{figure}
\begin{center}
\vspace{-0.45in}
\epsscale{1.5}
\plotone{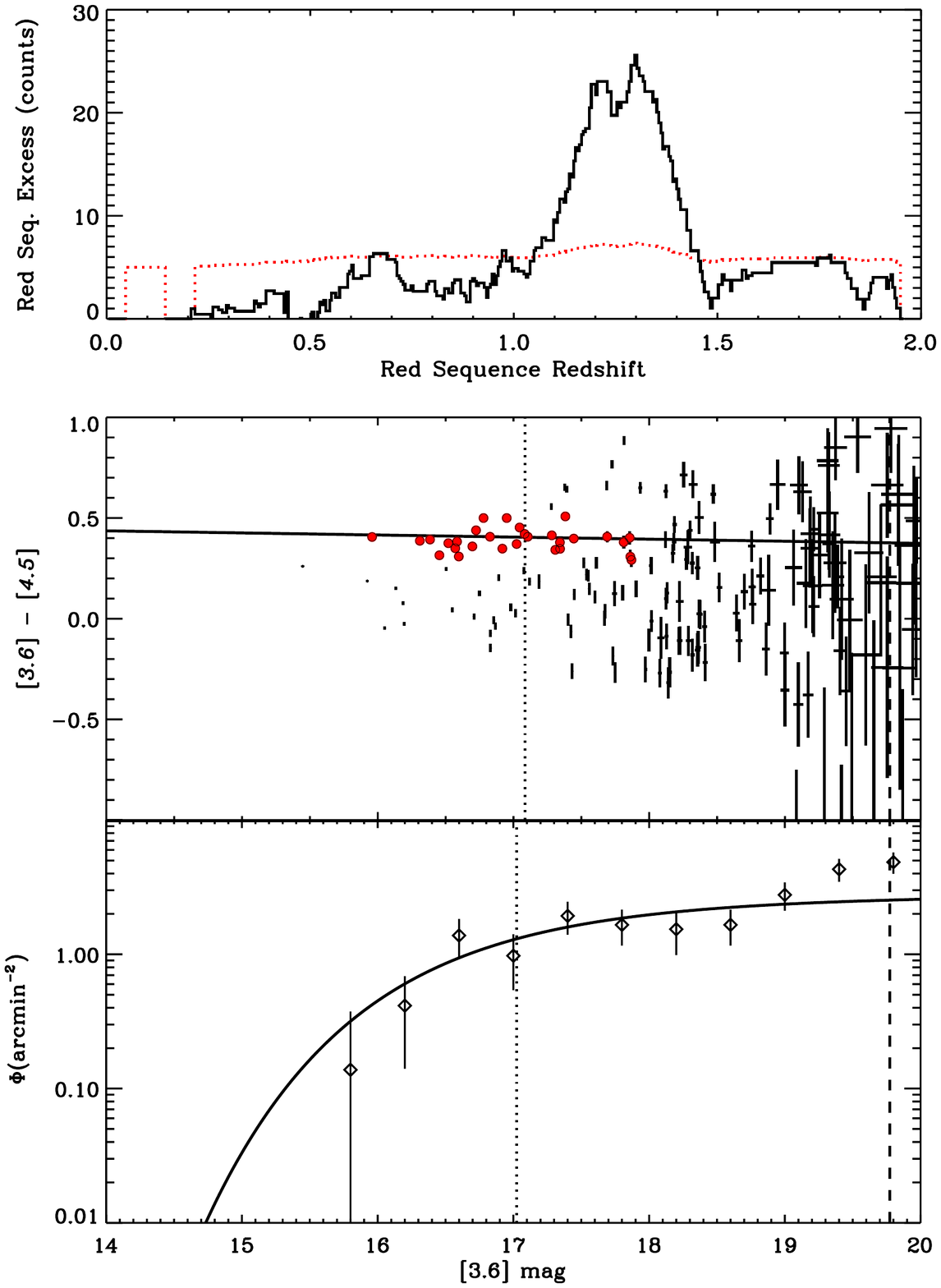}
\vspace{-0.35in}
\caption{
The cluster member finding for \cluster.  The top panel shows the excess of surface density of galaxies (above background) consistent within 2$\sigma$ of the SED model as a function of redshift.  The RMS level of the overdensity is shown in dotted red.  The peak overdensity is at \zpnoe\ at 3.5$\sigma$ above the background RMS.
The middle panel shows the color-magnitude diagram of all objects within 2 arcminutes of the SPT center coordinates with red-filled black circles showing the selected galaxies from the passively-evolving model at \zpnoe\ (black solid line) and brighter than m$^*+$1.  The inferred model m$^*$ is shown as a dotted vertical line and the [3.6] magnitude limit is shown as the dashed line.  
The bottom panel shows the Spitzer [3.6] galaxy luminosity function for \cluster.  The dotted vertical line shows the best fit m$^*$ in [3.6], and the dashed line shows the [3.6] magnitude limit.
}\label{f:optical}
\end{center}
\end{figure}

Infrared {\it Spitzer}/IRAC imaging was obtained in 2011 during Cycle 7
as part of a larger program to follow up clusters identified in the
SPT survey.   IRAC imaging is particularly important for the confirmation
and study of high-redshift SPT clusters such as \cluster\
where the optically faint members are strongly detected in the
infrared.   The on-target observations
consisted of $8\times100$\,s and $6\times30$\,s dithered exposures in bands
[3.6] and [4.5] to 10-$\sigma$ depths of 20.3 and 18.8 Vega magnitudes, respectively. The deep [3.6] observations
are sensitive to passively evolving cluster galaxies down to 0.1 $L^*$
at $z = 1.5$.  The data were reduced exactly as in \citet{brodwin10},
following the method of \citet{ashby09}.  Briefly, we correct for
column pulldown and residual image effects, mosaic the individual exposures, resample to
0\farcs86 pixels (half the solid angle of the native IRAC pixels), and
reject cosmic rays.

\subsection{Optical Spectroscopy}\label{ss:specobs}

Multislit spectroscopic observations were acquired for \cluster\ on the 6.5-meter Baade Magellan telescope
on UT 25-26 September 2011 using the f/2 camera on the IMACS spectrograph for a total integration time of 
11 hours.
The strategy and procedure were as described in \citet{brodwin10}, with the same
300 l/mm ``red'' grism and WB6300-9500 filter, but without the GISMO module (in order to increase throughput).
The galaxy target selection was based on the optical and infrared photometry, see Section \ref{s:results}.
Twenty-two 30-minute exposures were made in excellent to moderately-good seeing (0.4-0.7$^{\prime\prime}$) using one slit mask.  The resolution of the observations, as measured from the sky lines, was $5.2$\AA.  In a procedure identical to \citet{ruel12}, 
the COSMOS reduction package was used for standard CCD processing,
resulting in wavelength-calibrated 2D spectra. The 1D spectra were
then extracted from the sum of the reduced data.  Spectral features were identified by eye from inspection
of the 2D and 1D spectra, and redshifts were then obtained by using RVSAO routines.

\begin{figure*}
\begin{center}
\epsscale{1.20}
\vspace{-3.5in}
\plotone{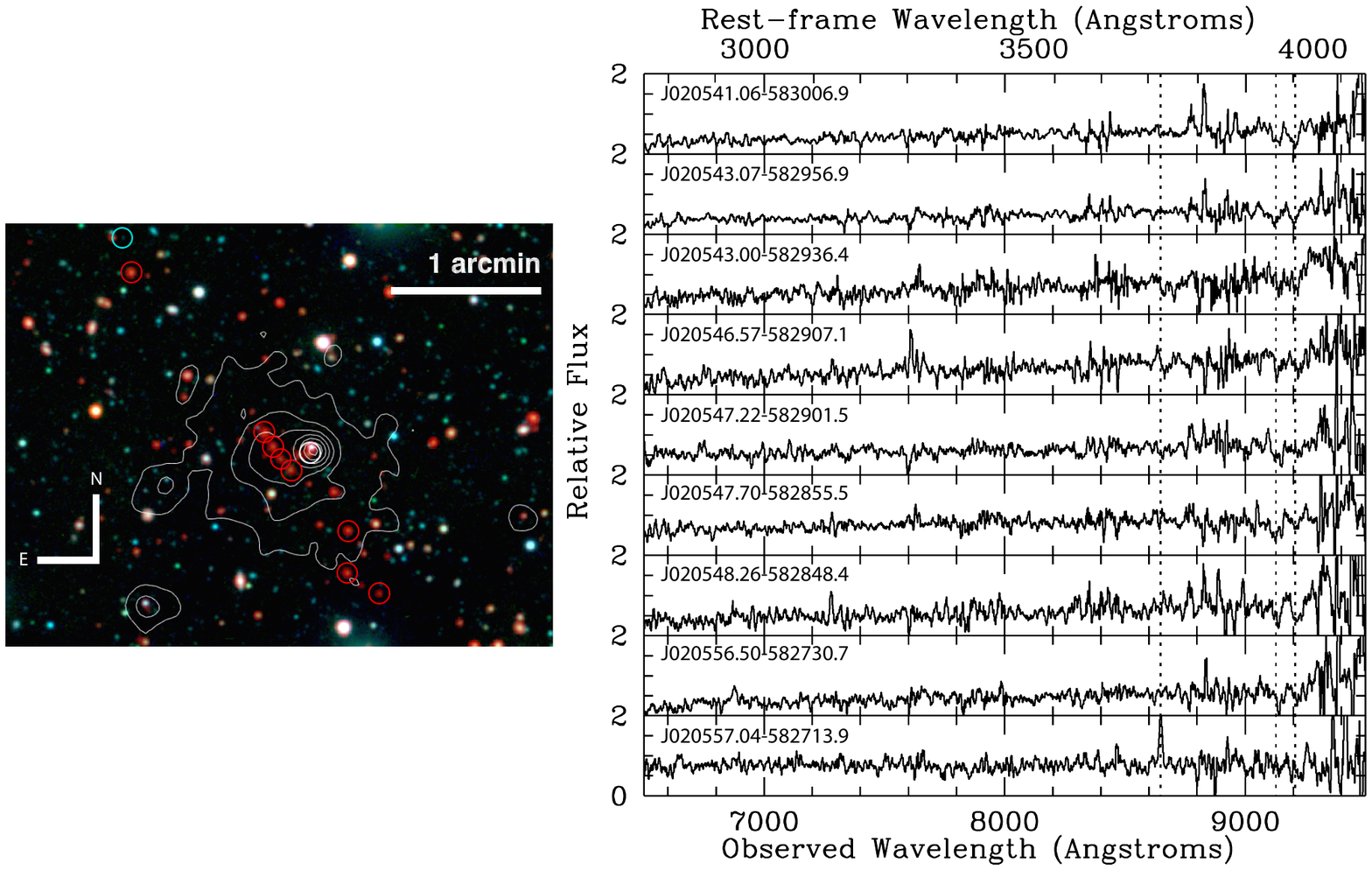}
\vspace{-1.7in}
\caption{
(Left)  Optical (r/i) and Spitzer/IRAC [3.6] image showing the galaxies confirmed by spectroscopy and
overlayed by the XMM-\emph{Newton} X-ray contours.  Cyan shows the [\ion{O}{2}] detection shown to the right (bottom panel), and red shows the members identified with Ca H\&K features shown to the right.  The white circle shows the X-ray point source position.  The frame subtends 4.5x3.4 arcminutes.  (Right)  Spectra of all 9 member galaxies with the [\ion{O}{2}] and Ca H\&K features indicated in vertical dotted lines.  Despite the long exposure time on Magellan, these features are faint due to the red color of passively galaxies.}\label{f:spec}

\end{center}
\end{figure*}

\subsection{X-ray Observations}\label{ss:xray}

A deep X-ray observation of \cluster\ was obtained by the
XMM-\emph{Newton} observatory (OBSID:0675010101) on UT 19-20 June 2011 
using the European Photon Imaging Camera (EPIC),
which consists of two Metal-Oxide-Silicon (MOS) arrays plus one fully
depleted p-n (PN) junction CCD array.  The total integration times were
 69~ks for the MOS arrays and 65~ks for the PN array.  The data
reduction and analysis were performed with \texttt{SAS~v11.0}
utilizing the XMM-\emph{Newton} Extended Source Analysis Software
package\footnote{\texttt{http://heasarc.gsfc.nasa.gov/docs/xmm/xmmhp\_xmmesas.html}}
\citep[XMM-ESAS, e.g.][]{snowden08}.  The net clean exposure time
is 57 and 39~ks in the MOS/PN arrays, respectively. Based on the
\citet{deluca04} diagnostics we find a $\sim30-40\%$ background
enhancement in the observation due to residual quiescent soft proton
contamination. The MOS2 CCD\#5 was in an anomalously high state and
we have removed it from further analysis.

We have also excised all point sources identified in the source
detection step. We have visually inspected the excision regions and
made conservative adjustments to their size. In particular, a point
source associated with a bright galaxy (bluer than the passively-evolving model) was
identified in the core region of the cluster ($\alpha$=02:05:45.4,
$\delta$=-58:28:58.3, $\sim12\arcsec$ west of the X-ray centroid)
and was removed with an excision radius of $\sim11\arcsec$ (see Figure \ref{f:spec}).


\section{Results}\label{s:results}

\subsection{Cluster Member Galaxies}

From the procedure described in S12, we measure a redshift based on the Spitzer IRAC photometry of \zp\ (see Figure \ref{f:optical}).  We fit a model of passively-evolved galaxies from \citet[][BC03]{bruzual03} to the data to determine the redshift.  The optical data were not deep enough to offer any additional constraint to the redshift except for the brightest of the cluster members (see section \ref{s:bcg}).  
The redshift estimator identified 32 galaxies with IRAC [3.6]-[4.5] colors consistent with this redshift (a 3.5$\sigma$ overdensity compared to the background), shown in Figures \ref{f:sz} and \ref{f:optical}.  From this list, we designed a multi-slit mask for the IMACS spectroscopic observations described in Section \ref{ss:specobs}, filling the mask with other targets, identified as galaxies with bluer colors relative to the model in the i-band and Spitzer data.

\subsection{Spectroscopy}\label{ss:specres}

Redshifts and other spectroscopic properties of member galaxies are listed in Table \ref{t:spec}.  Of the 47 slits designed into the mask, one spectroscopic member was identified from an [\ion{O}{2}] emission line, and 8 from Ca H\&K.  Figure \ref{f:spec} shows the spectra of the 9 cluster members.
The brightest cluster galaxy (BCG), defined as the brightest cluster member in [3.6], is at a redshift of \zbcg, and the combined robust (biweight) redshift of 9 cluster members is \zs.  We do not calculate the velocity dispersion due to the overwhelming intrinsic uncertainty in the derived mass estimates with $<$15 members \citep{saro12}.

We also estimate the star formation rate (SFR) for each cluster member from the integrated [\ion{O}{2}] flux which was corrected for galactic extinction (reddening) using the dust map from \citet{schlegel98} and scaled to match the i-band magnitude from IMACS imaging.  We do not correct for source dust extinction as we lack a well constrained NUV-Blue continuum measurement for most of these galaxies.  This is also consistent with our derived extinction from SED fits of the 4 brightest central galaxies (see Section \ref{s:bcg}).  We measured the continuum-subtracted flux centered on the [\ion{O}{2}] wavelength with a bin width of 8\AA\ (320~km/s) and converted to luminosity using the cluster redshift.  The SFR was estimated from the [\ion{O}{2}] luminosity using the scaling law from \citet{kennicutt98}.  The measured [\ion{O}{2}] flux and SFR (or 3$\sigma$ upper limits) are given in Table \ref{t:spec}.  

\begin{deluxetable*}{lccllccc}
\tabletypesize{\scriptsize}
\tablecaption{Spectroscopic members of
    SPT-CL J0205-5829 \label{t:spec}} \tablewidth{0pt} \tablehead{ \colhead{}
    & \colhead{R.A.} & \colhead{Dec.} & \colhead{} & \colhead{} &
    \colhead{Principal} & \colhead{[OII] flux\tablenotemark{b}} & \colhead{SFR\tablenotemark{b}} \\
    \colhead{ID} & \colhead{(J2000)} & \colhead{(J2000)} &
    \colhead{$z$} & \colhead{$\delta z$\tablenotemark{a}} &
    \colhead{Spectral Feature} & \colhead{($\times10^{-18}erg/cm^2/s/$\AA)} & \colhead{(\msun/yr)}} \startdata
J020556.50-582730.7 & 02:05:56.50 & -58:27:30.7 & 1.3219 & 0.0007 & Ca H\&K & 2.5$\pm$1.5 & 0.42$\pm$0.28 \\ 
J020548.26-582848.4\tablenotemark{c} & 02:05:48.27 & -58:28:48.4 & 1.3218 & 0.0005 & Ca H\&K & 1.4$\pm$0.9 & 0.24$\pm$0.16 \\ 
J020547.70-582855.5 & 02:05:47.70 & -58:28:55.6 & 1.3239 & 0.0003 & Ca H\&K & 1.2$\pm$0.5 & 0.19$\pm$0.11 \\ 
J020547.22-582901.5 & 02:05:47.23 & -58:29:01.6 & 1.3223 & 0.0005 & Ca H\&K & $<$6.7 & $<$1.40 \\ 
J020546.57-582907.1 & 02:05:46.57 & -58:29:07.1 & 1.3230 & 0.0005\tablenotemark{d} & Ca H\&K & $<$4.0 & $<$0.80 \\ 
J020543.00-582936.4 & 02:05:43.00 & -58:29:36.4 & 1.3119 & 0.0005 & Ca H\&K & $<$3.4 & $<$0.63 \\ 
J020543.07-582956.9 & 02:05:43.08 & -58:29:56.9 & 1.3186 & 0.0002 & Ca H\&K & $<$1.6 & $<$0.28 \\ 
J020541.06-583006.9 & 02:05:41.07 & -58:30:07.0 & 1.3210 & 0.0007 & Ca H\&K & $<$5.3 & $<$0.90 \\ 
J020557.04-582713.9 & 02:05:57.04 & -58:27:13.9 & 1.3106 & 0.0007 & [\ion{O}{2}] & 14.8$\pm$4.6 & 2.5$\pm$1.1 \\ 
\enddata
\tablenotetext{a}{Redshift errors are twice those given by RVSAO.}
\tablenotetext{b}{Based on integrated [OII] flux within 4\AA\ of the line peak with no source extinction correction.  Upper limits are 3$\sigma$.}
\tablenotetext{c}{BCG.}
\tablenotetext{d}{The RVSAO error was unphysically small and was adjusted up to a typical value.}
\end{deluxetable*}

\subsection{Brightest Central Galaxies}\label{s:bcg}

\begin{figure*}
\begin{center}
\vspace{-5.3in}
\epsscale{1.2}
\plotone{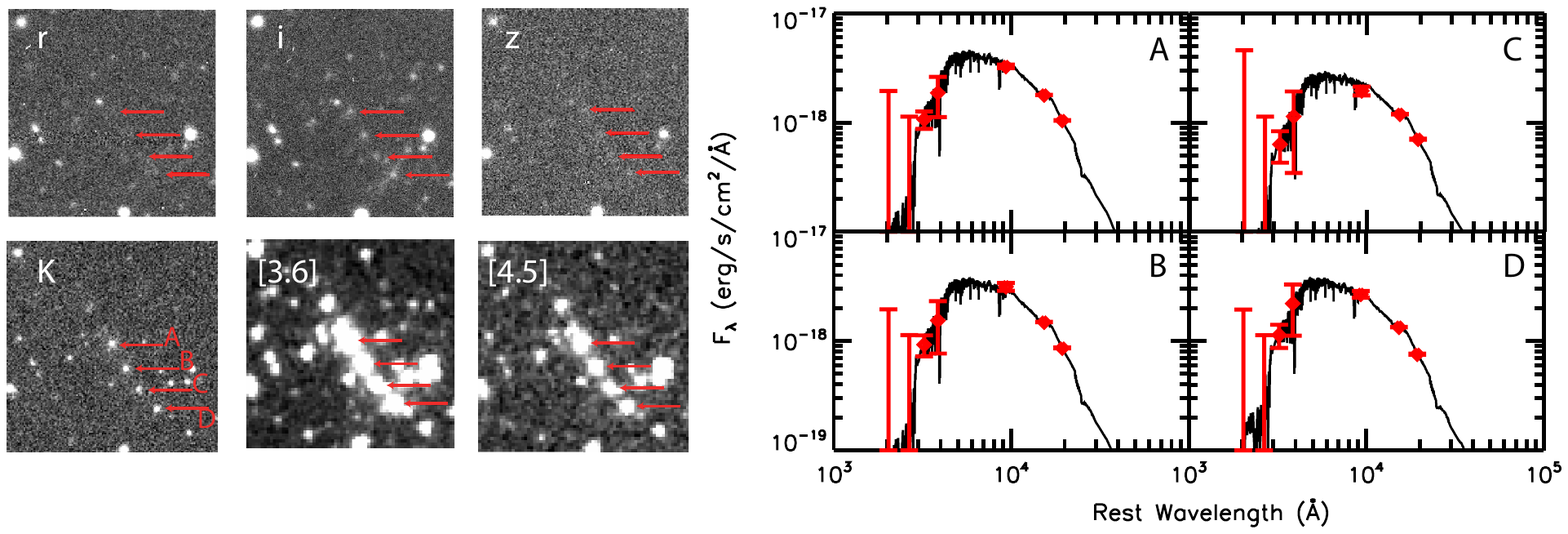}
\vspace{-0.5in}
\caption{
Best BC03 models for the four brightest galaxies in the central region of \cluster.  The left panel shows thumbnails of 6 observed filters (r, i, z, K$_s$, [3.6], and [4.5]) in the central $\sim$30$\arcsec$ from the brightest galaxy (A), with the four galaxies labeled in the K$_S$ image.  The right panels show the measured photometry in red with the best fit BC03 model for each overplotted.
}\label{f:bcg}
\end{center}
\end{figure*}

We selected the brightest central galaxies to be the four brightest galaxies consistent with the [3.6]-[4.5] model, within 1 arcminute of the SZ center.  We then use an analysis similar to the \citet{rosati09} spectral energy distribution (SED) fitting procedure.  To constrain the star formation history (SFH) of each of these galaxies, we fit an exponential-burst stellar population SED model at solar metallicity and Chabrier IMF from BC03 to the available photometry (see Figure \ref{f:bcg}), including magnitude lower limits, fix the redshift at \zsnoe, and add a source reddening model from  \citet{calzetti00}.
From the fit parameters, we calculate the rest-frame K-band luminosity, stellar mass (and corresponding stellar mass-to-light ratios), and age.  The uncertainties for these parameters are from the $\chi^2$ 68\% confidence intervals in the multi-dimensional sampled grid and checked by bootstrapping this procedure hundreds of times and were in good agreement.  
These parameters are presented in Table \ref{t:bcg}.  We find that all models give well-constrained K-band luminosities, mainly because the observed Spitzer [4.5] filter corresponds to 2 $\mu$m in the rest frame.  
The 4000\AA\ breaks are mainly constrained by the deep IMACS i-band measurement.

The rest-frame K-band luminosity of the brightest galaxy, at L $\sim4\times10^{11}$\lsun, is typical for BCGs in similar-sized clusters at $z<$0.25 based on previous X-ray \citep{haarsma10} or optical cluster studies \citep{lin04b,popesso07,brough08}, and smaller studies extending to higher redshifts ($z<$1) by \citet{whiley08}.  The derived stellar mass is also consistent with other studies of BCGs from X-ray samples at similar cluster masses and redshifts \citep{stott10}.  The derived ages from the BC03 model fits listed in Table \ref{t:bcg} suggest that the stellar mass of these brightest galaxies had formed by the time of the observed epoch was probably complete by redshift 2 or 3, or perhaps earlier, also consistent with previous studies of stellar ages of cluster galaxies at high redshift \citep{collins09,henry10}.  However this does not rule out the scenario of ``dry merging'' hierarchical build up of these galaxies between redshift 1.3 and 3.


\begin{deluxetable*}{c@{ }c@{ }c@{ }c@{ }c@{ }c@{ }c@{ }c@{ }}
\tabletypesize{\scriptsize}
\tablewidth{0pt}
\tablecaption{Brightest Central Galaxy Parameters\label{t:bcg}}
\tablehead{
\colhead{Galaxy} &
\colhead{$M_\mathrm{stellar}$\tablenotemark{a}} &
\colhead{M$_K$\tablenotemark{b}} &
\colhead{L$_K$\tablenotemark{c}} &
\colhead{M/L\tablenotemark{d}} &
\colhead{age\tablenotemark{e}} &
\colhead{$\tau$\tablenotemark{f}} &
\colhead{A$_V$\tablenotemark{g}}\\
\colhead{} &
\colhead{(10$^{11}$\msun)} &
\colhead{(Vega)} &
\colhead{(10$^{11}$\lsun)} &
\colhead{(\msun/\lsun)} &
\colhead{(Gyr)} &
\colhead{(Gyr)} &
\colhead{}}
\startdata
A       & 3.5$\pm$0.5 & -25.51$\pm$0.05 & 3.3$\pm$0.2 & 1.1$\pm$0.2 & 4.5$\pm$0.5 & 0.1$\pm$0.1 & 0.0$\pm$0.1 \\
B       & 2.9$\pm$0.5 & -25.31$\pm$0.05 & 2.7$\pm$0.2 & 1.1$\pm$0.2 & 4.5$\pm$0.8 & 0.1$\pm$0.1 & 0.0$\pm$0.1 \\
C       & 2.3$\pm$0.4 & -25.07$\pm$0.12& 2.2$\pm$0.3 & 1.1$\pm$0.4 & 4.5$\pm$0.6 & 0.1$\pm$0.2 & 0.1$\pm$0.2 \\
D       & 1.9$\pm$0.3 & -25.19$\pm$0.08 & 2.4$\pm$0.2 & 0.8$\pm$0.2 & 2.8$\pm$0.9 & 0.1$\pm$0.2 & 0.0$\pm$0.1 \\
\enddata
\tablenotetext{a}{Model Initial Stellar Mass at t=t$_\mathrm{form}$}
\tablenotetext{b}{Absolute rest-frame K Magnitude at t=t$_\mathrm{obs}$}
\tablenotetext{c}{Rest-frame K Luminosity at t=t$_\mathrm{obs}$}
\tablenotetext{d}{Stellar Mass to K Luminosity ratio}
\tablenotetext{e}{t$_\mathrm{obs}$-t$_\mathrm{form}$}
\tablenotetext{f}{e-folding timescale for SFR$\propto \exp(\mathrm{t}/\tau)$}
\tablenotetext{g}{Rest-frame V-band extinction in magnitudes}
\end{deluxetable*}

\subsection{NIR Luminosity Function}

As a further check on the cluster galaxy properties, we measure the observed [3.6]
(roughly rest H-band) luminosity function of galaxies with [3.6]-[4.5]
colors consistent with the BC03 model from our initial redshift estimate.
Galaxies selected as cluster members are within 1 Mpc (physical distance)
of the SZ-derived center and have [3.6]-[4.5] colors within 2$\sigma$ (based on each galaxy's photometric uncertainty) of the BC03 model.
We then measure the number density in 0.4 magnitude bins from the brightest cluster galaxy to 1 magnitude brighter than
the measured 10$\sigma$ magnitude limit (to reduce any systematic errors due to incompleteness in the catalog).
Field galaxy contamination was corrected by measuring the same quantity outside of the 1 Mpc aperture and subtracting.
We used the Schechter luminosity function,

\begin{equation}
\scriptsize
\Phi(m)=0.4ln(10)\Phi^*10^{-0.4\mu(\alpha+1)}exp(-10^{-0.4\mu}),
\normalsize
\end{equation}

\noindent{}where $\mu=m-m^*$ and allowed $\Phi^*$, $\alpha$, and m$^*$ to vary.  The final derived parameters and uncertainties 
are from the least squares fit to the data and bootstrapping the whole procedure thousands of times from the catalog selection stage.  We found the [3.6] best fit parameters are
$\Phi^*$=2.73 $\pm$ 0.31 arcsec$^{-2}$, $\alpha$=-1.02 $\pm$ 0.11 and m$^*$=16.58 $\pm$ 0.29 (Vega), which are roughly consistent with our previous model assumptions of $\alpha$=-1.0 and m$^*$=17.09 at this redshift, calculated from the evolving stellar population BC03 models, normalized to the Coma cluster luminosity function (see H10 for a discussion).  
Recent measurements of the luminosity function in evolved $z > 1$ clusters find a similarly flat faint-end slope, $\alpha$ \citep{mancone12}.  In contrast, less evolved high-redshift clusters have a paucity of faint galaxies, indicated by a shallower faint-end slope \citep[e.g.][]{rudnick12,lemaux12}.

This best fit luminosity function also corresponds to a richness measurement of N$_{gal}$=47$\pm$4 using the H10 procedure (integrating the luminosity function down to m$^*$+1 within a 1 Mpc physical radius of the BCG) and is consistent with the H10 sample of SPT-SZ clusters which are drawn from the same SPT-SZ significance although sampled at a different wavelength (observed i-band).

\subsection{SZ Mass Estimate}
We use an SZ mass estimate as described in R12 and \citet{benson11}, which is calculated from the 
Markov chain Monte Carlo (MCMC) method using available CMB, BAO, SNe, and SPT$_{CL}$ (from the R12 cluster sample) data.  The masses reported are posterior estimates based on the probability density function using the $\xi$ and redshift for \cluster, marginalized over uncertainties in the SZ and X-ray ($Y_X$) observable-mass scaling relations and cosmology.   
In Table \ref{t:mass}, we quote mass estimates with and without a Bayesian prior assumption on the underlying population of clusters. The expected bias on the flat-prior mass estimate is related to Eddington bias and affects the \cluster\ mass estimate at the $\sim10\%$ level. This bias is due to the steeply falling mass function which makes it more likely for \cluster\ to be a lower mass cluster that scattered up, than a higher mass cluster scattering down. The total uncertainty in mass ($\sim20\%$) is dominated a combination of the intrinsic scatter and the uncertainty in the normalization of the SZ-mass scaling relation.
In Table \ref{t:mass}, the mass estimates are given as $M_{500}$, defined as the mass within a radius in which the cluster has a density 500 times the critical density of the universe.  We can convert between this $M_{500}$ and $M_{200}$ with respect to $\rho_{mean}$, defined as the mass within a radius in which the cluster has a density 200 times the mean density of the universe, by assuming an NFW profile \citep{navarro97} and the mass-concentration relation by \citet{duffy08}.  Using this conversion, the $M_{200}$ masses are a factor of $\sim$1.8 times larger, such that the unbiased SZ mass-estimate is $M_{200}$=(\masssztwo)$\times10^{14} h_{70}^{-1} \mathrm{M}_{\sun}$

\subsection{X-ray spectroscopy with XMM-Newton}

We estimate the X-ray physical parameters of \cluster\ using an
iterative process over the cluster radius.  We measure the core-excised 
X-ray temperature, $T_X$, within \rd, defined as the radius inside which the
mass density is higher than 500 times the critical density of the universe.  We 
iterate over values of \rd\ so that the measured $T_X$ maintains consistency with the 
$M-T$ relation from \citet[][V09]{vikhlinin09b}.

For each value of \rd, we extract spectra and redistribution and
ancillary response files.  We excise all detected point sources from both the source and
background regions as well as the central $r\le 0.15\,\rd$ cluster
core region.  Given the significant residual quiescent
contamination (Sect.~\ref{ss:xray}), we opt to use a local background
model in the fitting procedure.
For each camera, we subtract a background spectrum extracted from an annulus centered on the
cluster between $160\arcsec$ and $320\arcsec$ in radius.
These radii were selected based on the cumulative count rate profiles so
that the annulus is not contaminated by cluster emission while still
lying on the same MOS chips as the source. 
The total number of background-subtracted source
counts is $\sim\xcts$ for all three cameras.
We use Xspec v12.5 to fit the spectra with a MeKaL model
\citep{mewe85,kaastra92,liedahl95} using C-statistics on minimally
binned spectra (i.e., binning only channels to obtain $\ge 1$ counts/bin).
From this fit to the spectrum, we measure the X-ray temperature.

We then use the measured X-ray temperature and the redshift from optical spectroscopy 
to infer a \md\ mass from the V09 $M-T$ relation, which we also 
convert to a corresponding \rd\ value.  Given this new value of \rd, we iterate on this process 
until two successive \rd\ estimates
differ by $\le 2.5\arcsec$ (equal to the bin size of our X-ray
images).  This criterion was reached in four iterations and we have
verified that the final solution is independent of the initial
\rd\ value.  The r$_{500}$ radius is 710~kpc($\sim85\arcsec$) and
the excised core region (i.e., $0.15\rd$) has a
radius of $\sim13\arcsec$ (roughly twice the PSF FWHM).

\begin{figure}
\begin{center}
\epsscale{1.6}
\plotone{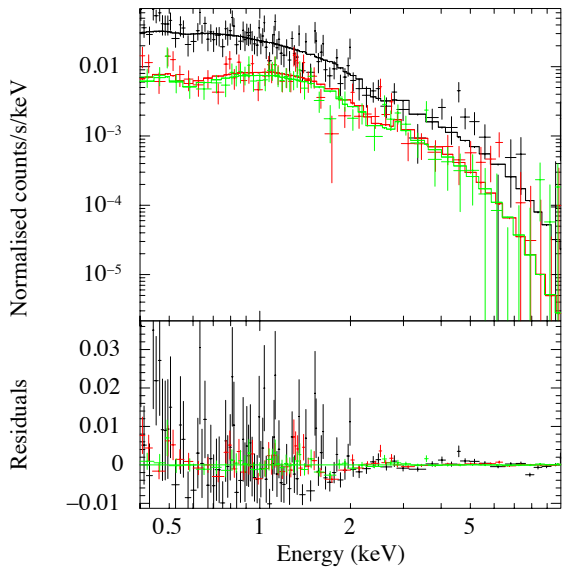}
\caption{
The X-ray spectrum of \cluster\ (black: PN,
  green: MOS1, red: MOS2).}\label{f:xray}
\end{center}
\end{figure}

The final spectrum is displayed in Figure
\ref{f:xray}. The best fit temperature is $T_X=$\xtempt~keV. This
corresponds to a mass of $M_{500}$=\masstxtmeas$h_{70}^{-1}$\msun\ using the relation from
V09.  There is no evidence that \cluster\ is unrelaxed from the X-ray morphology or the galaxy distributions, either on the sky or in the velocities, therefore we use the standard $M-T$ relation from V09 and make no corrections based on the dynamical state of the cluster.  In the $T_X$-based mass estimate,
we include the statistical uncertainty in the measurement of $T_X$, uncertainties in cosmology, and assume a 20\% intrinsic scatter in the $M-T$ relation, as noted by V09.

We constrain
the mean [Fe] abundance to Z=0.26$\pm$0.15Z$_\odot$ which is consistent with a typical
mean metallicity for the ICM in a massive galaxy cluster (Z$\sim$0.3Z$_\odot$) at
lower redshifts \citep[e.g.][]{matsumoto00,tozzi03,maughan08}.  The luminosity within \rd\ is \xlum\ in the rest frame.

We also note that we detect the Fe K line clearly in PN and MOS2 and
more weakly in MOS1. If we allow the redshift to vary
during fitting the best fit value from the joint fit is $z$=$1.39 \pm
0.02$, which is $\sim5\%$ larger than the optical spectroscopic redshift of \zs. The redshifts
derived from individual cameras are 1$\sigma$ consistent with the
joint PN+MOS1+MOS2 fit (except MOS1 which gives a slightly lower
redshift). This is one of the highest redshifts measured from
X-ray spectra \citep[cf.,][]{lloyd10}.

\begin{deluxetable}{l@{ }l@{ }l@{ }}
\tabletypesize{\scriptsize}
\tablewidth{0pt}
\tablecaption{Mass Estimates for \cluster\label{t:mass}}
\tablehead{
\colhead{Observable} &
\colhead{Measurement} &
\colhead{$M_{500} (10^{14} h_{70}^{-1} \mathrm{M}_{\sun})$}}

\startdata

{\bf SZ} $\mathbf{\xi}$       & 10.5                   & $\mathbf{\massszfive}$ \\
SZ $\xi$ (flat prior)         & 10.5                   & $5.2 \pm 1.1$ \\
$T_{X}$                       & \xtempt\ keV           & \masstxtfive \\
\hline\\
{\bf Combined}                & \nodata                & $\mathbf{\massbestshortfive}$

\enddata

\tablecomments{Unbolded masses indicate Eddington-biased mass
estimates, calculated using flat priors on mass.  Note that the SZ
(untargeted) flat-prior measurement suffers from a considerably different
Eddington bias than the $T_{x}$ (targeted) flat-prior estimate.
The $T_X$ mass estimate and the unbiased SZ $\xi$ mass estimate 
were used to generate the combined
mass estimate based on their probability distributions. (see Section \ref{s:combined}).}
\end{deluxetable}

\subsection{Combined Mass Estimate}\label{s:combined}

We follow \cite{foley11} and calculate a joint estimate using the SZ mass and X-ray mass estimates of \cluster.  We assume the uncertainties are uncorrelated between the two masses.  This allows for a more straightforward evaluation of the posterior probability distribution function (PDF),

\begin{equation}
P(M|\xi,T_X) \propto P(M) P(\xi|M) P(T_X|M),
\end{equation}

\noindent{}where $P(M)$ is the Tinker halo mass function \citep{tinker08}, $P(\xi|M)$ is the flat-prior SZ mass estimate PDF, and $P(T_X|M)$ is the flat-prior $T_X$ mass estimate PDF.  
As calculated in Sections 3.5 and 3.6, we use the X-ray and SZ mass estimates derived from the observables Tx and $\xi$, respectively, which were marginalized over uncertainties in their scaling relations and cosmology.  
We find a combined, unbiased, mass estimate to be $M_{500}$=\massbestfive$h_{70}^{-1}$\msun.  Converting to $M_{200}$ as above gives $M_{200}$=\massbesttwo$h_{70}^{-1}$\msun.


\section{Discussion}\label{s:discussion}

The galaxy members in \cluster\ were identified via [3.6]-[4.5] color, and the significant overdensity of these is  how this cluster was initially confirmed after it was identified by the SZ effect.  The measured overdensity (richness) of these galaxies is consistent with other SPT-SZ clusters, a sample that has a median mass of $M_{500}\sim3.3\times10^{14}h_{70}^{-1}$\msun.  From SED fitting, the BCG and 3 other bright central galaxies have luminosities and stellar masses typical of central galaxies in clusters of similar mass at lower redshift and have derived stellar population ages greater than $\sim$3 Gyr.   This suggests that most of the eventual stellar mass in these galaxies are already present at $z=$1.3 and the vast majority of these stars were formed by $z\sim$3.  The actual assembly scenario of these galaxies cannot be constrained yet.

The quiescent SEDs and the amount of [\ion{O}{2}] in the spectra of the central galaxies suggest there is very little ongoing star formation ($<$0.5\msun/yr) in the center of \cluster, meaning there is no \emph{strong} cooling flow mechanism depositing new gas into these galaxies 
\citep{hu85,heckman89,crawford99,hatch07,mcdonald10}.   
However, it should be noted that only a very strong cooling core would be discernible in [\ion{O}{2}] \citep{mcdonald11,santos11} and a better indicator would be H$_\alpha$.  
The central X-ray point source noted in Section \ref{ss:xray} could possibly be a central AGN that is suppressing the star formation in the cluster.  Alternatively, given the age of the cluster and the X-ray cooling time, there might not have been sufficient time for a strong cooling flow to form.  Either interpretation is consistent with the general lack of star formation and strong cooling flows at $z > 0.5$ found in previous studies \citep{santos08,vikhlinin07,samuele11,mcdonald11}, where the number density of strong cooling flows increases dramatically at $z<$0.5 while AGN and merger activity decreases.  We do not see any indication of a major merger either in the X-ray morphology or galaxy distribution.

From the X-ray spectrum, we found that the cluster gas has a metallicity consistent, albeit with significant uncertainty, with massive clusters at lower redshift.  
This is also consistent with several studies \citep[see][for a review]{baldi12} of ICM metal abundances over a range of redshifts that found little or no evidence of evolution from $z<$1.4.
Several studies have found that this enrichment can happen over a timescale of 1 Gyr \citep{pipino04} and 
settle into the central region within a cluster crossing time (1 Gyr).  The best-fit metallicity of \cluster\ would suggest that the bulk of the metal production could have been completed by $z\sim$2.5, however more X-ray observations are needed to say this with high statistical significance.

The optical and infrared data have shown that the stellar populations of the most massive central galaxies are already well-evolved, suggesting that the assembly of these galaxies happened within the preceding 2-4 Gyr.  It has been suggested that this timeline may depend on the mass of the cluster, as there is some evidence that the BCGs at $z > 1$ in lower mass clusters have not fully assembled  \citep{stott10}.  It may become possible to see this change over the full SPT mass range and may probe different regimes where other feedback modes dominate.  
Such a study of a large SZ-selected sample has the potential to directly measure the build up of the stellar mass as a function of redshift and cluster mass.   

\subsection{Rarity}

Although \cluster\ was included in the sample used in R12 for cosmological analysis, 
we did not assign a goodness of fit to the model, 
so it is interesting to quantify the probability of having found this cluster in the full 2500\,\sqdeg SPT-SZ survey.  
We use the full 2500\,\sqdeg SPT-SZ survey area in order to avoid a posteriori selection of the area in which \cluster\ was found, which could artificially boost the apparent rarity.
We follow \citet{foley11} and compute the probability of finding a cluster at higher mass and higher redshift than \cluster.  We do so by sampling the cosmological and scaling relation constraints of the 
CMB+BBN+BAO+HST+SN+SPT$_{CL}$ chain from R12 
and producing a posterior statistical mass estimate $P(M|\xi,z)$ at each step in the chain.  We then compute the expected number of clusters at higher mass and higher redshift

\begin{equation}
  \tilde{x}_{>z>M} = \int_{0}^\infty \int_z^\infty \frac{dN}{dM' dz'} \int_{0}^{M'} P(M''|\xi, z) dM'' dz' dM',
\end{equation}
where $\frac{dN}{dM dz}$ is the mass function as calculated following \citet{tinker08}.  The median point in cosmological and scaling relation parameter space predicts $\tilde{x}_{>z>M} = 0.07$ clusters at higher mass and higher redshift than \cluster\ in 2500\,\sqdeg.  

However, as noted by \citet{hotchkiss11,hoyle12,waizmann12a,waizmann12b}, this statistic has a small expectation value due to the fact that it requires a cluster of simultaneously higher mass and higher redshift than a particular object. This statistic does not consider the fact that many similarly rare clusters could exist with a slightly higher mass and lower redshift or lower mass and higher redshift. Instead, we follow the treatment of \citet{hotchkiss11} and compute the probability of finding the particular value of $\tilde{x}_{>z>M} = 0.07$, corresponding to \cluster\  for an ensemble of simulated $2500\,{\rm deg}^2$ surveys.
We then create a normalized histogram of the resulting values of $\tilde{x}_{>z>M}$ for the rarest cluster in each catalog and integrate the area under the curve from 0 to the value of $\tilde{x}_{>z>M}$ for the particular cluster in question. This statistic, unlike $\tilde{x}_{>z>M}$ itself, has an expectation value of 0.5. We note that it depends only very weakly on the details of the simulation or the point in cosmological or scaling relation space at which the simulations are performed. This metric suggests that this cluster is not at all surprising with a probability of 0.69 of finding at least one cluster as rare as \cluster\ in 2500\,\sqdeg.

As a comparison to other more rare clusters in the SPT-SZ survey, using the same statistic we find 0.21 for \uber, which was considered in \citet{foley11} and 0.05 for \rarestcluster\ (ACT-CL 0102-4915), currently the rarest cluster in 2500\,\sqdeg SPT-SZ survey.



\section{Summary and Conclusions}\label{s:conc}

We report the massive galaxy cluster \cluster\ at \zsnoe\ discovered in the first 720~\sqdeg of the SPT-SZ survey and present results of follow up observations at optical, infrared, and X-ray wavelengths.  The galaxy population of this cluster shows a strong red sequence with a luminosity function consistent with that of lower redshift SZ-selected clusters.  Galaxy SED fits to an exponentially decaying SFR stellar population, the [Fe] abundance from the X-ray spectrum, and the lack of [\ion{O}{2}] emission in most of the optical galaxy spectra suggests that the bulk of the star formation happened at an earlier epoch ($z > 2.5$).  Optical spectroscopy of 9 galaxies confirms the cluster redshift at \zs, also roughly consistent with X-ray spectroscopy which gives $z$=$1.39 \pm0.02$.  This establishes \cluster\ as the highest-redshift SZ-selected galaxy cluster verified by spectroscopy, and the second most massive SZ-selected cluster known at $z > 1$.  Based on the X-ray temperature, \cluster\ is consistent with being more massive than XMM2235 at $z$=1.39 with $T_X=$\xtemptrosati\ from \citet{rosati09} (but the uncertainties in both temperatures are much larger than the measured difference).

The measured mass observables (from the SZ and X-ray temperature) are consistent and give a combined mass estimate of $M_{500} = \massbestfive h_{70}^{-1}$\msun.  Although not the most massive SZ-discovered cluster, it demonstrates that a cluster of this mass has enough time to form during the first 5 Gyr of the universe, and the existence of this rare object appears to be fully consistent with general expectations for a flat $\Lambda$CDM cosmological model.

In general, we find that \cluster\ has properties similar to clusters with the same mass at lower redshift.
This is extremely important in the context of an ultimate goal of an unbiased and low scatter mass-calibration of clusters over a wide range of redshifts for cosmological studies, and provides new insight to the assembly of the rarest and most massive structures in the universe.

\begin{acknowledgments}


\bigskip

The South Pole Telescope program is supported by the National Science
Foundation through grant ANT-0638937.  Partial support is also
provided by the NSF Physics Frontier Center grant PHY-0114422 to the
Kavli Institute of Cosmological Physics at the University of Chicago,
the Kavli Foundation, and the Gordon and Betty Moore Foundation.
Galaxy cluster research
at Harvard is supported by NSF grant AST-1009012.  Galaxy cluster
research at SAO is supported in part by NSF grants AST-1009649 and
MRI-0723073.  The McGill group acknowledges funding from the National
Sciences and Engineering Research Council of Canada, Canada Research
Chairs program, and the Canadian Institute for Advanced Research.
X-ray research at the CfA is supported through NASA Contract NAS
8-03060.  The Munich group acknowledges support from the Excellence
Cluster Universe and the DFG research program TR33.
This work is based in part on observations obtained with the Spitzer Space
Telescope (PID 60099), which is operated by the Jet Propulsion
Laboratory, California Institute of Technology under a contract with
NASA.  Support for this work was provided by NASA through an award
issued by JPL/Caltech.  Additional data were obtained with the 6.5~m
Magellan Telescopes located at the Las Campanas Observatory,
Chile and the Blanco 4~m Telescope at Cerro Tololo Interamerican
Observatories in Chile.
R.J.F.\ is supported by a Clay Fellowship.  B.A.B\ is supported by a KICP
Fellowship, M.Bautz acknowledges support from contract
2834-MIT-SAO-4018 from the Pennsylvania State University to the
Massachusetts Institute of Technology.  M.D.\ acknowledges support
from an Alfred P.\ Sloan Research Fellowship, W.F.\ and C.J.\
acknowledge support from the Smithsonian Institution.

\end{acknowledgments}

\bibliographystyle{fapj}
\bibliography{../../BIBTEX/spt.bib}


\end{document}